\documentclass[floatfix,
reprint,
superscriptaddress,
 amsmath,amssymb,
 aps, 
 pre,
]{revtex4-2}
\usepackage{graphicx,color}
\usepackage{dcolumn}
\usepackage{bm}
\usepackage{siunitx} 
\usepackage[normalem]{ulem}
\usepackage{hyperref}
\hypersetup{
    colorlinks=true,  
    linkcolor=blue,   
    citecolor=red,  
    urlcolor=cyan     
}

\def\d{{\mathrm{d}}}

\usepackage{lipsum}

\begin{document}


\title{\textbf{Impact of Initial Charge Distributions on the Kinetics of Charged Particle Coagulation} 
}%

\author{Gustavo Castillo}
 \affiliation{Instituto de Ciencias de la Ingeniería, Universidad de O'Higgins, Avenida Libertador Bernardo O'Higgins 611, Rancagua, Chile}
 \email{Contact author: gustavo.castillo@uoh.cl}

\author{Nicolás Mujica}
\affiliation{
 Departamento de Física, Facultad de Ciencias Físicas y Matemáticas, Universidad de Chile, Avenida Blanco Encalada 2008, Santiago, Chile
}%



\date{\today}

\begin{abstract}
We investigate the kinetics of particle aggregation within the framework of the Smoluchowski coagulation equation, extending it to account for electrostatic interactions among charged clusters. Using a stochastic Monte Carlo implementation, we examine how different charge distributions and net system charge affect cluster growth dynamics. 
Electrostatic interactions are incorporated directly into the classical Brownian collision kernel, yielding charge-dependent modifications of the collision rates that may either enhance or suppress aggregation depending on the signs and magnitudes of the interacting charges.
Our simulations reveal distinct regimes of growth: at intermediate times, charge heterogeneity accelerates or delays aggregation depending on the initial underlying charge distribution, while at long times the system tends toward quasi--stationary states whose properties depend on the net charge. Comparisons between Gaussian and Cauchy--Lorentz initial charge statistics highlight the role of heavy-tailed distributions in promoting faster cluster growth. These findings contribute to a unified understanding of coagulation kinetics in charged particulate systems, with potential implications for aerosol and astrophysical coagulation processes, volcanic ash aggregation, and clustering in industrial fluidized granular beds.
\end{abstract}

\maketitle


\section{Introduction}

One of the fundamental issues in aerosol and granular physics is clustering and coagulation \cite{goldhirsch1991clustering-410,Puglisi1998,Friedlander2000,Tinsley2000,Royer2009}. It is a complex problem, with various interaction forces and multiple scales in play simultaneously. Indeed, far from simple volumetric repulsion, the dynamics of aerosol and granular systems can be driven by a variety of interactions: van der Waals forces, magnetic forces, electrostatic Coulomb and polarization forces, viscous drag, and, in the presence of even minute amounts of liquid, capillary bridges or ice coatings (which can strongly increase stickiness), and also dissipative collisions. More generally, how these forces interact and drive a solid particle or droplet system to clustering and coagulation is important in many natural and industrial settings. Some examples are: meteorite~\cite{schrader2018retention-597} and planet formation \cite{Steinpilz2020,wurm2021understanding-e3a}, dust, ice and volcanic ash aggregation in the atmosphere and also under extreme conditions \cite{Tinsley2000,Brown2012,Cameron2025}, electrically induced raindrop enlargement and rainfall rate intensification \cite{Guo2021,Mudiar2021,Mudiar2022}, particle assembly under arid conditions \cite{Shinbrot2006}, electrostatic clustering in particle-laden turbulent flows \cite{ruan2024effects-7fc}, pharmaceutical aerosol deposition in human airways \cite{Karner2011,Xi2014,Bessler2023},
particle clumping during coffee grinding~\cite{harper2024moisture-controlled-5d0}, clustering in fluidized industrial beds \cite{cocco2010particle-172,xu2011visualization-5fd,Wang2020}, and for efficient water treatment with negligible generation of by-product wastes \cite{Moussa2017}. 

For all of the examples above, electrostatic interactions are essential to the assembly or coagulation process.  However, the origin of these electric charges has not been fully resolved. What we do know is that in many of the described situations they arise from contact electrification, also known as tribocharging: two colliding or sliding surfaces will exchange electric charges; when separated, their surface charges will be modified. The fact that this can occur for \textit{identical} materials during contact is even more puzzling. Issues as basic as the nature of the charge carrier are also not understood, and several candidates have been proposed, including electrons \cite{lowell1986triboelectrification-cf7}, ions \cite{mccarty2008electrostatic-331} or nanoscale pieces of charged material \cite{baytekin2011mosaic-182}.  In order to model how charges affect clustering processes, we should know how charges are exchanged at contacts and how these exchanges lead to a stable distribution of charges; but these are still open questions. The ``patch” model for charging has shown some success 
in addressing many contact electrification phenomena \cite{Apodaca2010, Harris2019, Grosjean2020,Grosjean2023b,Grosshans2025}, in particular involving identical materials. However, there is no strong evidence for \textit{what} these patches actually are~\cite{Grosjean2023}. 

Concerning charge distributions, these are very difficult to obtain accurately in natural settings. Geophysicists chasing naturally charged granular media, e.g., dust storms \cite{Sow2011} or volcanic plumes \cite{Gilbert1991}, can gather some rough charge data, but not enough for reproducible distributions. For astrophysical situations, experiments to measure charges of dust particles or pebbles in a protoplanetary disk (PPD) are currently impossible. For atmospheric aerosols, e.g. $\si{\micro\meter}$ drops in clouds, little data is available and the charge distribution is assumed to be Gaussian \cite{Guo2021}. One approach is to measure charge probability distribution functions (PDFs) in the laboratory. Indeed, experiments with sub-$\si{\milli\meter}$ oxide particles have shown that a large ensemble of same-material, initially neutral same-size grains, which undergoes many particle contacts/collisions, will reach a stationary and stable charge distribution, with both positive and negative charges. This has been well-documented experimentally through videography measurements of particle trajectories under an imposed electric field in microgravity conditions \cite{Waitukaitis2013,Waitukaitis2014,Lee2015,Steinpilz2020,Mujica2023,Lara2026}. These experiments have also shown the efficiency of electrostatic-induced clustering \cite{Lee2015,Steinpilz2020} and the charge polarity segregation induced by grain size effects \cite{Waitukaitis2014}. The measured PDFs are non-Gaussian with fat tails. When measurement uncertainties are considered carefully, the charge PDF of monodisperse oxide particles can be determined very accurately; it is close to a Cauchy--Lorentz distribution~\cite{Mujica2023,Lara2026}. Other experiments measure individual charges using Faraday cups; grains collide with a plate or repeatedly with the inner walls of a tube, revealing charge distributions with non-Gaussian tails \cite{Haeberle2018}. Charge distributions for different volcanic ash ensembles have also been measured using Faraday cups, with some evidence of non-Gaussian tails \cite{MendezHarper2016}. They have also been measured for \si{\micro\meter} size lycoperdon particles charged by shaking, i.e. by contact electrification, also revealing a distribution with fat tails \cite{Emets1991}. In general, the existence of heavy tails is important because they imply large numbers of highly-charged particles, which strongly affect how particles coagulate \cite{cocco2010particle-172,teiser2025growth-abd,Steinpilz2020,Lee2015}. Additionally, they may bear relevance to the underlying mechanism for charge transfer \cite{Haeberle2018, Lee2018, Grosjean2023}, as the non-Gaussian element indicates the existence of memory effects: charge exchanges depend on particle charges before contact~\cite{Matsuyama1995,Matsusaka2010,Han2021}. All this experimental evidence indicates that non-Gaussian, heavy-tail distributions are ubiquitous to particulate systems that charge through contact electrification. 

In this work, we focus on the case of coalescence of a few large objects starting from a monodisperse gas of electrically charged particles under Brownian motion. This fixes the size range from $\si{\nano\meter}$ to a few $\si{\micro\meter}$ for aerosols in terrestrial settings \cite{Andreotti_Forterre_Pouliquen_2013}, and up to $\SI{100}{\micro\meter}$ for dust grains that are strongly coupled to the dense gas in a PPD \cite{blum2004grain}. The main question that drives this research is: what is the effect, in cluster size and growth rate, of the initial charge distribution in the formation of clusters? We use the Smoluchowski coagulation equation framework with collision rates that include Coulomb electrostatic interactions. We compare Gaussian and Cauchy--Lorentz initial charge distributions to highlight the role of heavy-tailed distributions, for both globally neutral and non-neutral charged systems. As our focus is the effect of the initial charge distribution, we do not consider charge exchanges during the subsequent system's evolution. We demonstrate that a monodisperse system of charged particles with a heavy-tailed distribution will naturally form large clusters much faster than a system with an initial Gaussian distribution. Moreover, because large clusters in heavy–tailed charge distributions should have a high electrostatic interaction energy, they are expected to exhibit enhanced internal cohesion and increased resistance to external perturbations. This has been indeed observed at a larger size scale for charged granular matter~\cite{Lee2015,teiser2025growth-abd}.

Our results should be relevant for a variety of electrically charged particle systems, in particular aerosols in terrestrial settings and dust in a PPD. The former are known to charge in some cases by contact electrification \cite{Friedlander2000,Karner2011}, and non-Gaussian distributions with heavy tails have been measured \cite{Emets1991}. Concerning the latter, after the first stages of growth by sticking, at the $\si{\micro\meter}$ scale, we expect that larger compact clusters, between $10-\SI{100}{\micro\meter}$, will exchange charges by contact electrification as well \cite{Schwaak2024,Becker2024,Onyeagusi2025}. Thus, their growth in this range and beyond should be more efficient due to the existence of very high charged particles. Our findings may also be important for the aggregation of particles in volcanic ash plumes \cite{Brown2012,MendezHarper2016,Cameron2025} and industrial fluidized granular beds \cite{cocco2010particle-172,xu2011visualization-5fd}; although these are not driven by Brownian motion, but by a background turbulent flow instead, in both systems charges are induced by contact electrification, which by its memory effects should lead to non-Gaussian distributions. 

Finally, polarization effects are important in these systems, because the same sign-charged particles that are close enough experience an attractive electrostatic force \cite{Cameron2025,Lee2018,Nakajima1999,Lindgren2018}. This is another mechanism of enhancement for electrostatic induced coagulation \cite{Guo2021,Cameron2025}, which is not considered in the kernel of our model. In this sense, our result can be considered as a “lower bound" for faster and strongly cohesive granular cluster growth. 

This paper is organized as follows: in section \ref{SectII} we layout the theoretical framework of the Smoluchowski coagulation equation with electric charges, and in section \ref{SectIII} we describe the numerical methods. Finally, in sections \ref{SectIV} and \ref{SectV} we present our results and conclusions, respectively.

\section{Theoretical Framework}
\label{SectII}
  \subsection{Smoluchowski coagulation equation with charges}
  For particles with concentrations $n_i$ and $n_j$, the collision frequency of particles of mass $m_i$ and $m_j$ and electric charge $q_i$ and $q_j$ immersed in a gas is defined as
$N_{ij} = \beta(m_i,m_j, q_i, q_j)\, n_i n_j,$
where $\beta(m_i,m_j, q_i, q_j)$ is the collision kernel, determined by the particle masses, electric charges, as well as by gas properties such as temperature, pressure, and flow conditions. The evolution equation for the number density of clusters of size $k$ is determined by the Smoluchowski equation~\cite{smoluchowski1916drei}:
\begin{eqnarray}
    \frac{\d n_k}{\d t}&=&\frac{1}{2}\sum_{i+j=k}N_{ij}-\sum_{i=1}N_{ik}\\
    \frac{\d n_k}{\d t}&=&\frac{1}{2}\sum_{i=1}^{k-1}\beta(m_i,m_k-m_i, q_i, q_j-q_i)n_i n_{k-i}\nonumber\\
    & &-n_k\sum_{i=1}^\infty \beta(m_i,m_k,q_i,q_k)n_i,\label{eq:smol}
\end{eqnarray}
where the first term accounts for the formation of clusters of size $k$ by binary aggregation, and the second term accounts for their loss due to further coagulation. The solution depends on the form of $\beta_{ij}=\beta(m_i,m_j,q_j,q_j)$, which is determined by the collision mechanism of particles~\cite{friedlander2000smoke}. The Smoluchowski equation is formulated under the assumption of spatial homogeneity, with no correlations in the positions of clusters, and permits coagulation between any pair of clusters regardless of their sizes. Also, Eqn.~(\ref{eq:smol}) conserves the number of particles and the global electric charge. 

In the absence of charges, analytical tractability is limited to a few idealized, non-physical kernels~\cite{wattis2006introduction-fda}. Conversely, the presence of electric charges has been shown to enhance aggregation and to promote the onset of gelation—a transition characterized by the emergence of clusters with masses far exceeding those of the initial particles, leading to an apparent breakdown of mass conservation~\cite{ivlev2002coagulation-885}.  
\section{Methods}
\label{SectIII}
  \subsection{Direct Simulation Monte Carlo (DSMC)}
  In this work, the Smoluchowski equation is solved using the direct simulation Monte Carlo (DSMC) method, originally proposed by Kruis {\it et al.} for coagulation~\cite{Kruis2000}, and later extended to incorporate nucleation and surface growth~\cite{Maisels2004}. The method is based on recasting collision rates into interaction probabilities between particle pairs. At each time step, the colliding pair and the corresponding inter-collision time are selected stochastically according to the collision frequencies $\beta_{ij}$.
  
  Since only coagulation is considered, the total number of clusters in the system decreases monotonically, which in turn reduces the statistical accuracy of the simulations and renders the method progressively less reliable. To mitigate this limitation, we adopt the top-up strategy introduced in~\cite{Liffman1992}. In this approach, the system is duplicated once the particle concentration has decreased by half, effectively doubling both the simulation volume $V$ and the number of particles $N$. In this way, the particle concentration is preserved, ensuring that the underlying physics of the system remains unchanged while maintaining adequate statistical resolution.
  
In order to study the role played by the initial charge distribution on the evolution of the system, we explore two distinct cases. At $t=0$, the system is composed of $N=1\times10^4$ monomers with an electric charge that follows either a Gaussian or a Cauchy--Lorentz distribution. Also, in order to examine the impact of the width of the initial distributions, we use the concept of interquartile range $IQR$, defined as the difference between the 75th and 25th percentiles, instead of the usual standard deviation, which is not well defined for the Cauchy--Lorentz distribution. In our simulations, we use normalized units for the charge so that $q_i\to q_i/\sqrt{2\pi\varepsilon d_0 k_B T}$, where $d_0$ is the monomer diameter. Lengths are measured in units of $d_0$, and time is measured in units of the Brownian
coagulation time $t_0 = 3\eta d_0^3/(2k_B T)$. With this choice, the prefactor $2k_B T/(3\eta)$ in the Smoluchowski kernel becomes unity in the nondimensional formulation.
In the Monte Carlo implementation, however, collision events are sampled within a finite system of $N$ particles. To ensure that the total event rate remains $\mathcal{O}(1)$ and properly scaled with system size, the kernel is further normalized by a factor $1/N$. Therefore, the factor $1/N = 10^{-4}$ appearing in the simulations reflects the stochastic normalization of the algorithm rather than the nondimensionalization of the Smoluchowski equation. We vary the initial interquartile range among the values $IQR_0 =\{1,5,10,50,100,500\}$. Finally, we also compare the evolution of the system depending on its net charge. Thus, we define the parameter $\lambda$ as a way to measure the net charge through the relation $\lambda =\sum_i q_i/IQR_0$. We explore the values $\lambda=\{0,0.01,0.1\}$. The initial charges are generated according to the following procedure:  
We first draw $N/2$ charges $q_i$ and give the remaining $N/2$ particles 
their exact opposites $-q_i$. 
An overall offset is then applied so that each charge is shifted 
according to $q_i \rightarrow q_i + \lambda$. This construction ensures  that the total charge of the system 
vanishes exactly (within numerical precision) in the neutral  case ($\lambda = 0$). The set of parameters explored is shown in Table~\ref{tab:params}.
  \begin{table}[t!]
      \centering
      \begin{tabular}{l|c}
      \hline
      \hline
      Size Distribution& Monodisperse\\ 
          Charge Distribution & Gaussian, Cauchy--Lorentz\\
          Inital interquartile range ($IQR_0$)& $1,5,10,50,100, 500$\\
          Net charge parameter $\lambda$& $0,0.01,0.1$\\
          Number of particles $N$& 10000\\
          Number of time-steps&60000
      \end{tabular}
      \caption{Parameters used in the DSMC simulations.}
      \label{tab:params}
  \end{table}
  All the results presented here correspond to the average of 10 different realizations with the same set of parameters.
\subsection{Kernel}
It is well known that when particles diffuse due to Brownian motion in a gas they cluster together in a process called Diffusion-limited aggregation (DLA)~\cite{witten1981diffusion-limited-261}. It has also been proven that when Coulombian interaction is taken into account, the collision rate for particles of volume $v_i$ and $v_j$, and charge $q_i$ and $q_j$ is given by~\cite{zebel1958zur-d39,fuchs1965mechanics-6c5,lianze2005analytical-cde}
\begin{align}
    \beta_{ij} &= \frac{2k_BT}{3\eta}\left(\frac{1}{v_i^{1/3}}+\frac{1}{v_j^{1/3}}\right)\left(v_i^{1/3}+v_j^{1/3}\right)f_{ij},
    \end{align}
with
    \begin{align}
    f_{ij} &= \frac{\kappa_{ij}}{\exp(\kappa_{ij})-1},\\
    \kappa_{ij}&=\frac{q_iq_j}{2\pi\varepsilon(v_i^{1/3}+v_j^{1/3})k_B T}.
\end{align}
In the previous expressions, $k_B$ is the Boltzmann constant, $T$ is the temperature, and $\eta$ and $\varepsilon$ are the fluid viscosity and the electric permittivity of the gas, respectively. The exponent $1/3$ comes from assuming that the volume of the agglomerates scales with a characteristic radius $a$ as $v\propto a^3$ (all the results we report are with fractal dimension $D_f =3$. It is worth to mention that we observe no qualitative differences in the evolution of the system with other values of $D_f$). $f_{ij}$ accounts for the effect of the electrostatic interaction in the collision rates. $\kappa_{ij}$ denotes the ratio of the electrostatic potential energy to the thermal energy $k_B T$. In the limiting case of uncharged particles ($\kappa_{ij}=0$), the correction factor $f_{ij}$ equals unity and the collision kernel reduces to the \emph{field-free} Smoluchowski form. When the particles carry opposite charges, $\kappa_{ij}$ is negative and the correction factor remains positive but smaller than one, as can be seen from a series expansion of the exponential.
This implies that collisions occur more rapidly than in the uncharged case.
Conversely, for like-charged particles, $\kappa_{ij}$ is positive and the correction factor exceeds unity, leading to a reduced collision rate relative to neutral particles.\\

    \subsection{Observables and definitions}
    \begin{figure*}[ht!]
    \centering
    \includegraphics[ width=\linewidth]{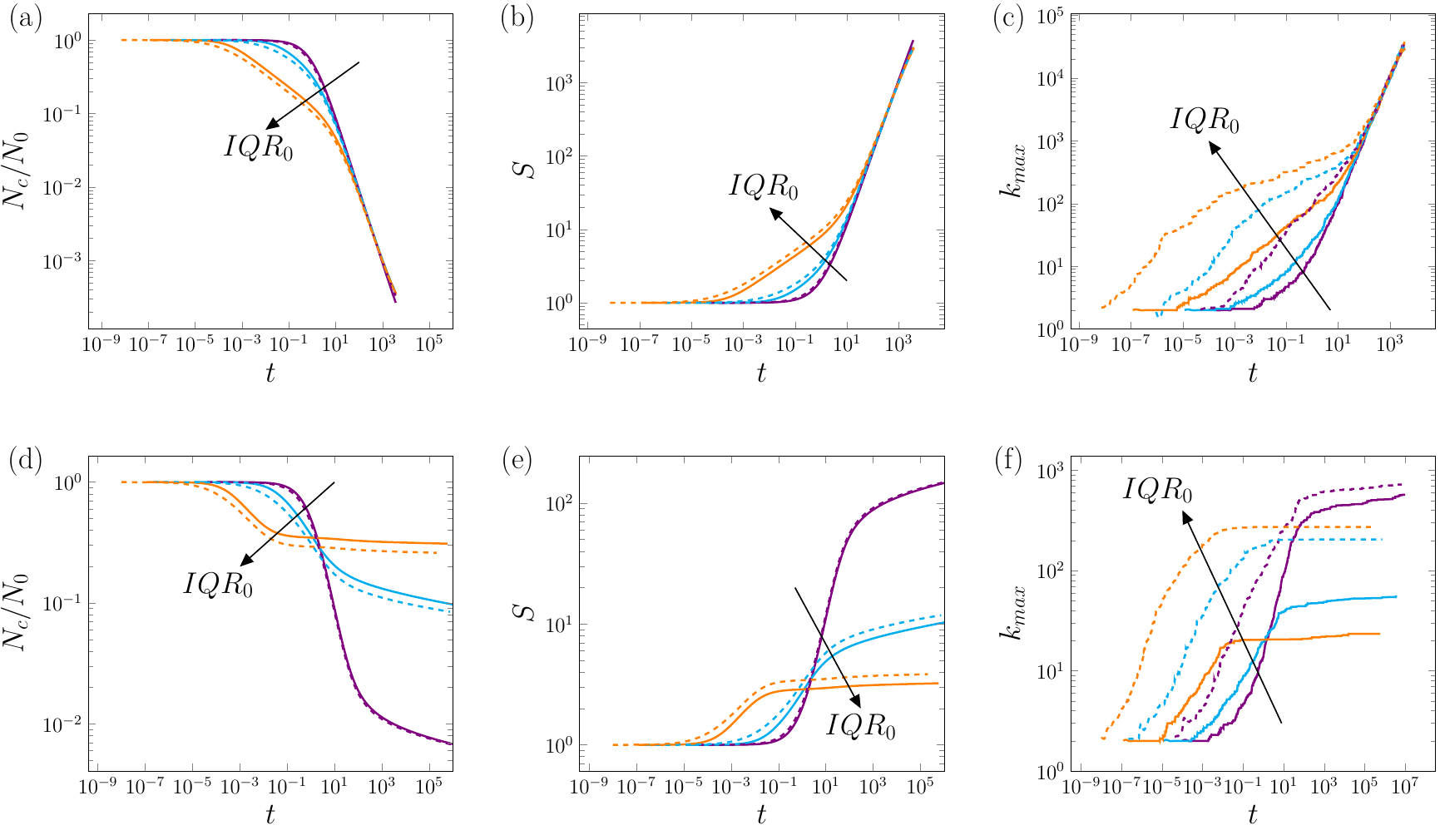}
    \caption{Time evolution of global quantities for neutral $\lambda =0$ (top row) and non-neutral $\lambda= 0.1$ (bottom row) systems. Panels show (a,d) normalized number of clusters, $N_c/N_0$, (b,e) mean cluster size, $S$, and (c,f) largest cluster size, $k_{max}$. In each panel, results are compared for Gaussian (continuous) and Cauchy--Lorentz (dashed) initial charge distributions. At intermediate times, cluster growth is significantly faster for the Cauchy--Lorentz case, particularly evident in the non-neutral system. For neutral systems, at long times all cases converge toward the self-preserving distribution (SPD), confirming asymptotic universality. Results are shown for different initial interquartile ranges $IQR_0 = \{1, 10, 100\}$.}
    \label{fig:fig1}
\end{figure*}

\paragraph{Number of clusters and mean cluster size.}We define $N_k(t)$ as the number of clusters containing $k$ monomers
($k$--mers) at time $t$, and $p_k(t)$ as the probability that a randomly
chosen cluster has size $k$.
The total number of clusters and the mean cluster size are then given by
\begin{align}
N_c(t) &= \sum_{k=1}^{\infty} N_k(t), \\
S(t) &= \sum_{k=1}^{\infty} k\,p_k(t)
     = \frac{\sum_{k=1}^{\infty} k\,N_k(t)}{\sum_{k=1}^{\infty} N_k(t)}.
\end{align}

\paragraph{Tail Index.}

Because the Cauchy--Lorentz distribution possess heavier tails than the Gaussian, we quantify the behavior of the extreme regions of the charge distributions using the tail index.
This parameter characterizes the ``heaviness'' or ``thinness'' of the tails of a probability distribution~\cite{crovella1999estimating-dc3}. Specifically, a distribution is said to have a tail index $\gamma$ if, 
for large $x$, its probability density decays as a power law 
$x^{-(\gamma+1)}$. 
Thus, the Cauchy--Lorentz distribution has $\gamma = 1$, whereas the 
Gaussian, whose tails decay faster than any power law, is formally 
characterized by $\gamma = \infty$. 
To estimate this parameter from our data we employ the 
\emph{Hill tail-index estimator}, denoted $\gamma_H$, chosen for its 
simplicity of implementation and its proven suitability for heavy-tailed 
samples~\cite{hill1975simple-2e9}. The estimator is applied to the largest $m$ order statistics of each data set, corresponding to the upper (right) tail of the distribution and to values exceeding a data-dependent threshold. 
In the present analysis, we choose $m = \lfloor 0.1\,N \rceil$, such that the tail index is estimated from the largest \SI{10}{\percent} of the data. 
This choice lies within a range where the Hill estimator exhibits a stable plateau, and all results were verified to be robust with respect to moderate variations of $m$ around this value. It is worth noting, however, that when applied to data drawn from a Gaussian distribution the Hill estimator systematically underestimates the tail index, yielding a finite value of approximately $\gamma_H \approx 4$,  despite the true value being infinite.

\section{Results}
\label{SectIV}
\subsection{Evolution of global quantities}
As a means to globally characterize how the system evolves, we study the number of clusters $N_c$, the mean cluster size $S$ and the largest cluster size $k_{max}$ as a function of time. 

\paragraph{Number of clusters.}

Figures \ref{fig:fig1}(a) and (d) show the number of clusters as a function of time for different values of the initial interquartile range $IQR_0$, comparing both neutral and charged systems.
Overall, there is no significant difference between the Gaussian and Cauchy--Lorentz charge distributions.
However, at intermediate times (around $t\sim10^{-2}$) the number of clusters decreases more rapidly as $IQR_0$ increases.
At later times, in the neutral case, the system approaches a self-preserving distribution (SPD), displaying a power-law decay in $N_c$.
For the neutral case, this SPD solution is independent of $IQR_0$, whereas in the charged (non-neutral) case the asymptotic distribution retains a clear dependence on $IQR_0$.

\paragraph{Mean cluster size.}
Something analogous is observed in the mean cluster size, Figs.~\ref{fig:fig1}(b) and (e). At intermediate times, the cluster growth is appreciably faster for wider initial distributions. At later times, {in the non-neutral case}, when clusters grow to sizes for which the Coulomb barrier becomes comparable to or larger than the thermal energy 
($U_c \gtrsim k_B T$), the coagulation rate for collisions between large clusters is strongly suppressed. This suppression breaks the scale-free growth and introduces an effective cutoff $k_c$, so that for $k \gg k_c$ the growth becomes extremely slow or practically arrested. This feature has been previously reported in systems with different net charge~\cite{dammer2004self-focusing-e28}.

\begin{figure}[ht!]
    \centering
    \includegraphics[width=0.95\linewidth]{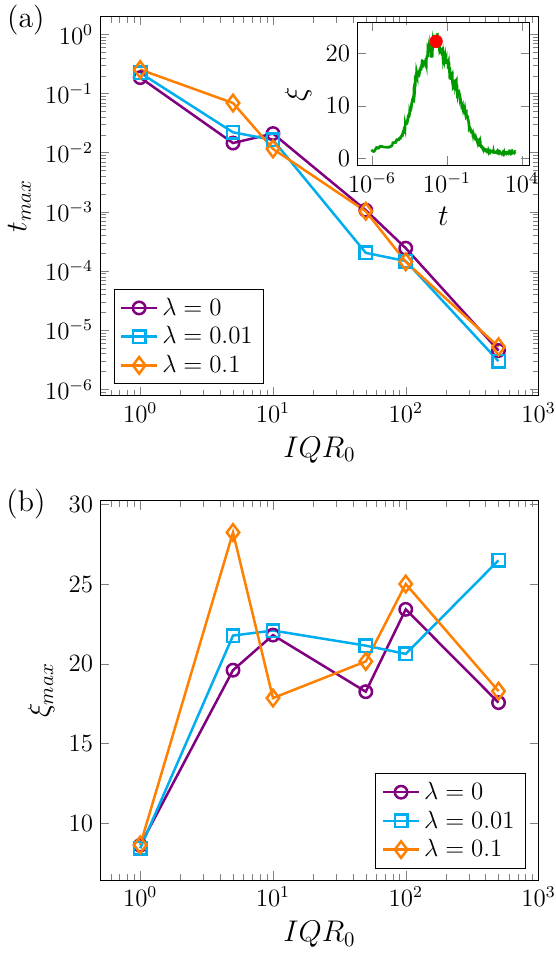}
    \caption{Panel (a) illustrates the characteristic time $t_{\mathrm{max}}$ at which the ratio $\xi \equiv k^{\mathrm{cauchy}}_{\max}/k^{\mathrm{gauss}}_{\max}$ reaches its maximum, while panel (b) presents the corresponding maximum value $\xi_{\max}$. Results are presented for different values of the net charge parameter $\lambda$. The inset displays the full temporal evolution of $\xi(t)$ for a representative example, with $\lambda = 0$ and $IQR_0=10$. The red dot marks the maximum obtained after smoothing the curve with a moving average procedure.}
    \label{fig:fig2}
\end{figure}

\paragraph{Largest cluster size.}
In Figs.~\ref{fig:fig1}(c) and (f) we present the evolution of the largest cluster size, $k_{max}$, as a function of time for the neutral and  charged cases, respectively. A clear distinction emerges between the  Gaussian and Cauchy--Lorentz initial charge distributions: the differences  are most pronounced at intermediate times. At long times, the neutral system approaches the previously discussed SPD, whereas in the non-neutral case the dynamics converge to a quasi--stationary state, with the Cauchy--Lorentz distribution attaining a larger asymptotic value of $k_{\max}$ than the Gaussian case as $t \to \infty$.
To quantify these differences, we introduce the ratio $\xi \equiv k_{max}^{\mathrm{Cauchy}} / k_{max}^{\mathrm{Gauss}}$ and plot it as a function of time (see Fig.~\ref{fig:fig2}) for several values of the initial interquartile  range $IQR_0$. We find that the time at which $\xi$ attains its maximum  decreases with increasing $IQR_0$, while the maximum value itself grows with $IQR_0$ and eventually saturates around  $\xi_{max} \approx 20$, largely independent of the net charge. In other words, a nearly twenty–fold enhancement of the largest cluster mass for the Cauchy--Lorentz distribution relative to the Gaussian case. 

The average size of the largest cluster can be evaluated for each case at time $t_{max}$. Its mass is $M = k_{max} m_0$, where $m_0$ is the monomer mass. For $D_f = 3$, the largest cluster diameter scales as $D = k_{max}^{1/3} d_0$. Using the data reported in Figs.~\ref{fig:fig1}(c) and \ref{fig:fig1}(f) we obtain $D^{\rm Cauchy}/D^{\rm Gauss} \approx 2-3$ over the range of $IQR_0$ and $\lambda$ considered. Although this diameter increase may appear
modest, the time $t^*$ required for the Gaussian case to reach the same cluster size is dramatically larger: for the neutral system $t^*/t_{max}$ is of order $20$, $800$, and $20000$ for $IQR_0 = 1$, $10$, and $100$, respectively; in the non-neutral case,  for $IQR_0\geq10$, $k_{max}$ is never reached in the total time of our simulation in the case of an initial Gaussian charge distribution.

  \begin{figure*}
    \centering
    \includegraphics[width=0.67\linewidth]{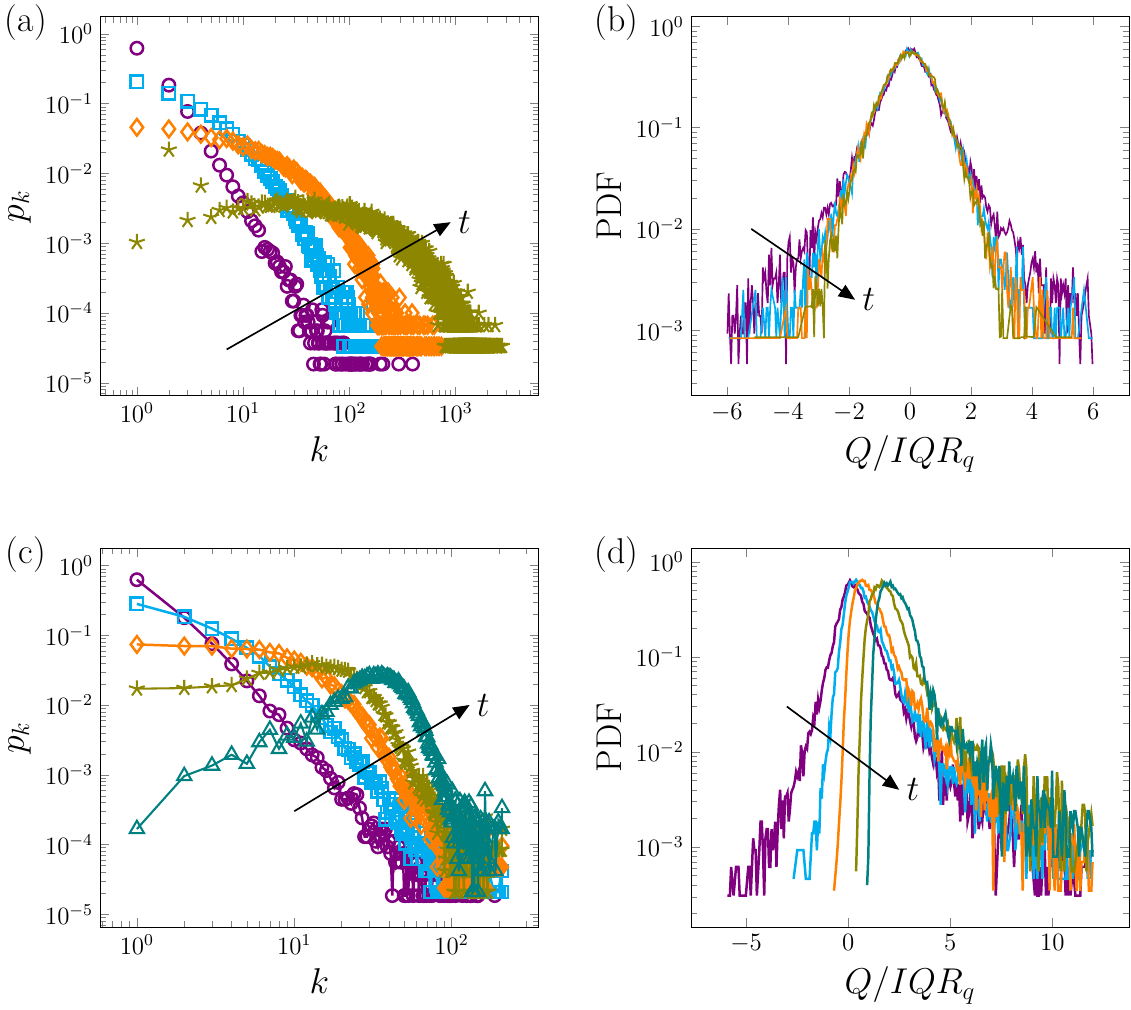}
    \caption{Cluster size distributions (left) and Probability Density Functions (right) of cluster charges at representative times for neutral (top row) and non-neutral (bottom row) systems ($\lambda = 0.1$). In (b) and (d), the charge is normalized by the instantaneous interquartile range $IQR_q$. Within each panel, results are shown for Cauchy--Lorentz initial charge distributions (the Gaussian case is qualitatively similar). At long times, for the neutral case, all cases collapse onto the universal SPD, independent of initial charge distribution or $IQR_0$. For the non-neutral case, the entire distribution progressively shifts toward larger charge values as the system evolves.}
    \label{fig:fig3}
\end{figure*}

  \subsection{Cluster size and charge distributions}
In Figs.~\ref{fig:fig3}(a) and (c) we present the cluster size distribution 
$p_k$ for systems initialized with a Cauchy--Lorentz charge distribution, 
considering both the neutral and charged cases. 
As expected for the neutral system, the probability of small clusters 
progressively decreases while that of large clusters increases as the 
system evolves, reflecting the scale–free nature of the aggregation 
process. 
In contrast, when a net charge is present, $p_k$ develops a pronounced 
peak, indicating the emergence of a characteristic cluster size and 
signaling a clear departure from the scale–free behavior of the neutral 
system. 
Regarding the charge distribution [Figs.~\ref{fig:fig3}(b) and (d)], the neutral case remains symmetric, 
as expected, and its tails gradually become less heavy over time, 
approaching a Gaussian profile. 
For the charged case, a marked asymmetry is evident, and the entire 
distribution progressively shifts toward larger charge values as the 
system evolves. 
These two coupled trends reveal the physical mechanism at work in the non-neutral
charged system: the increasing electrostatic repulsion not only limits 
the growth of the largest aggregates but also biases the population 
toward highly charged clusters of intermediate size, producing a 
self–regulated state in which size and charge coevolve.

  \subsection{Charge statistics across clusters}
\begin{figure*}
    \centering
    \includegraphics[width=0.67\linewidth]{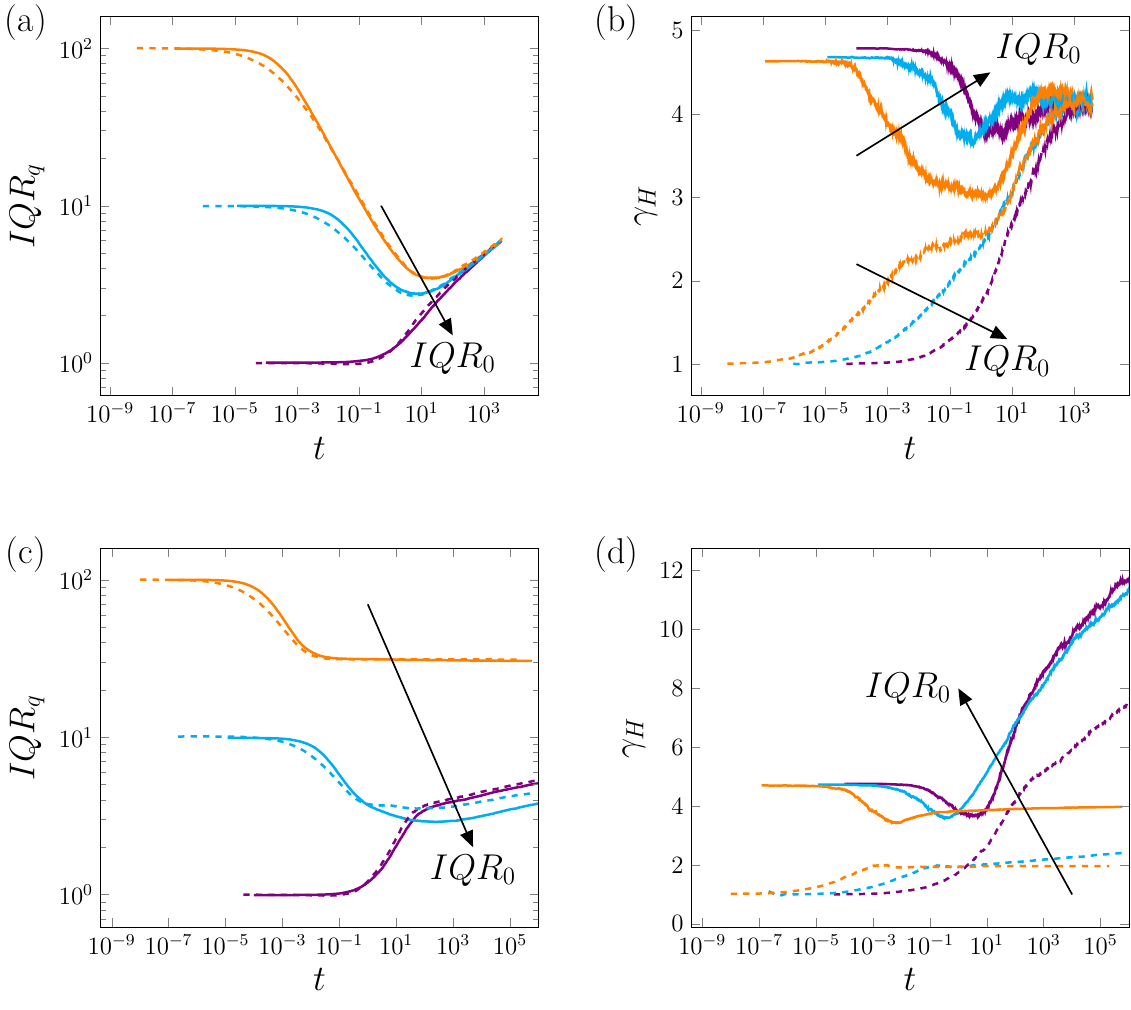}
    \caption{Time evolution of charge distribution statistics for neutral (top row) and non-neutral (bottom row) systems: (a,c) instantaneous interquartile range, $IQR_q$, and (b,d) Hill estimator of the tail index, $\gamma_H$. Gaussian (continuous lines) and Cauchy--Lorentz (dashed lines) initializations are compared in each panel. Results are shown for different initial interquartile ranges $IQR_0 = \{1, 10, 100\}$.}
    \label{fig:fig4}
\end{figure*}

     \paragraph{Interquartile range.} 
In Figs.~\ref{fig:fig4}(a) and (c) we show the instantaneous interquartile range ($IQR_q$) 
of the charge distribution as a function of time for both the neutral and  charged cases. 
For the neutral system, after a short transient all curves converge at $t \sim 10^2$ to the same asymptotic width variation with time, independently of the initial value $IQR_0$. 
This collapse reflects the fact that, in the absence of a net charge, charge fluctuations are progressively averaged out by aggregation, so the distribution naturally relaxes to a universal Gaussian‐like width. The charged case behaves differently: at long times, the instantaneous interquartile range, $IQR_q$,  approaches a constant value, but in this case the final width retains a clear dependence on the initial $IQR_0$. This sensitivity indicates that the persistent net charge limits the degree of statistical mixing; electrostatic repulsion slows down the exchange of charges between clusters, allowing memory of the initial spread to survive even after the system has reached its asymptotic regime.

    \paragraph{Tail index (Hill estimator)}
To probe the behavior of the distribution tails, we computed the tail index using Hill’s estimator, as shown in Figs.~\ref{fig:fig4}(b) and (d). 
We first examine the evolution of the Hill estimator in the neutral case.
Although the neutral system ultimately converges toward a universal
Gaussian–like charge distribution, its temporal evolution exhibits a
structured, multi–stage behavior that reflects the gradual loss of memory
of the initial charge statistics.
At early times, the dynamics are \emph{memory–dominated}: the estimated
tail index closely mirrors the form of the initial charge distribution,
with Gaussian initial conditions yielding $\gamma_H \sim 4$ and
Cauchy--Lorentz initializations producing significantly smaller values,
indicative of heavy tails.
In this regime, both the initial tail shape and the initial width $IQR_0$
strongly influence the statistics, as aggregation has not yet proceeded
far enough to redistribute charge efficiently.

At intermediate times, the neutral system enters a \emph{crossover
regime} in which charge mixing becomes increasingly effective.
Frequent collisions between oppositely charged clusters promote partial
charge cancellation, leading to a rapid reduction of extreme charge
fluctuations.
This process manifests itself as a nonmonotonic evolution of the Hill
estimator, including a pronounced minimum whose depth and temporal
location depend on $IQR_0$.
During this stage, the initial width controls how quickly the heavy tails
are eroded, while the distinction between Gaussian and Cauchy--Lorentz
initial conditions progressively diminishes.

At late times, all neutral systems reach a \emph{Gaussian–dominated
regime}, characterized by the saturation of the Hill estimator at
$\gamma_H \approx 4$, independently of both the initial tail shape and
$IQR_0$.
This convergence reflects the self–averaging nature of charge aggregation
in the absence of a net charge: repeated mergers act as a random–walk
process in charge space, driving the distribution toward a universal
Gaussian form and completely erasing memory of the initial charge
statistics.

The non–neutral case displays a markedly different behavior.
While an initial memory–dominated regime is again observed, the subsequent
evolution does not lead to a universal Gaussian attractor.
Instead, three distinct stages emerge in which the roles of the initial
tail shape and width are fundamentally different.
For narrow initial charge distributions, Gaussian and Cauchy--Lorentz
initializations converge toward distinct asymptotic values of the Hill
index, indicating that the system retains a measurable imprint of the
initial tail shape even at long times.
For broader initial conditions, the Hill estimator increases steadily with
time, signaling the progressive suppression of extreme charges by
electrostatic repulsion.\\
This late–time behavior corresponds to a \emph{Coulomb–dominated regime},
in which charge and cluster size become strongly correlated and extreme
charge fluctuations are selectively pruned.
Importantly, the unbounded growth of $\gamma_H$ in this regime does not
indicate heavier power–law tails; rather, it reflects the effective
thinning of the charge distribution as Coulomb barriers inhibit the
formation and survival of highly charged aggregates.
As a result, non–neutral systems do not evolve toward a single universal
charge statistics, but instead display a continuum of asymptotic behaviors
that depend sensitively on both the initial tail shape and the initial
width of the charge distribution.

\paragraph{Charge per particle.}
A complementary characterization of the charge statistics is provided by
the average charge per particle, $Q/k$, which quantifies how the net
cluster charge scales with its mass.
In the neutral system, this quantity decreases monotonically with time and
vanishes asymptotically, as expected from self–averaging: while the total
charge of a cluster grows only as $\sqrt{k}$ due to the central limit
theorem (CLT), its mass increases linearly with $k$, yielding
$|Q|/k \sim k^{-1/2} \to 0$ as $k$ increases.

For net–charged systems, the situation is fundamentally different.
In our model, each monomer is initially assigned a charge drawn
independently from a fixed parent distribution with mean $\mu_0$ and
variance $\sigma_0^2$, and clusters form exclusively through aggregation.
As a result, the charge of a cluster of size $k$ is simply the sum of the
charges of its $k$ constituent monomers, $
Q = \sum_{i=1}^{k} q_i .$
Under these assumptions, the charge statistics of a cluster of fixed size
$k$ follow directly from the central limit theorem:
\begin{equation}
    \mathbb{E}[Q\,|\,k] = k\,\mu_0,
\qquad
\mathrm{Var}(Q\,|\,k) = k\,\sigma_0^2 .
\end{equation}
Consequently, the conditional distribution of the charge per particle is approximately
Gaussian,
\begin{equation}
P\!\left(\frac{Q}{k}\,\Big|\,k\right)
\;\approx\;
\mathcal{N}\!\left(\mu_0,\;\frac{\sigma_0^2}{k}\right).
\end{equation}
so that $Q/k \to \mu_0$ as $k$ grows and relative fluctuations decay as
$k^{-1/2}$.
We emphasize that this Gaussian behavior applies \emph{only} to the charge
statistics conditioned on a fixed cluster size.
The unconditional charge distribution, obtained by integrating over the
cluster size distribution, corresponds to a size–weighted mixture of
Gaussians and need not be Gaussian when strong correlations between charge
and cluster size persist.\\
This convergence reflects the fact that the present model does not include
explicit charge–exchange mechanisms between particles: once monomer charges
are assigned initially, the charge of each cluster is entirely determined
by its aggregation history.
As a result, any evolution of the charge statistics arises solely from
aggregation and size–charge correlations, allowing us to isolate the role
played by the initial charge distribution.
Consistent with the trends observed in the interquartile range $IQR_q$ and
the tail–index analysis, neutral systems therefore lose memory of their
initial charge statistics, whereas net–charged systems retain a finite
mean charge density and a persistent imprint of their initial conditions.

  \subsection{Joint size--charge structure}
\begin{figure*}
    \centering
    \includegraphics[width=0.67\linewidth]{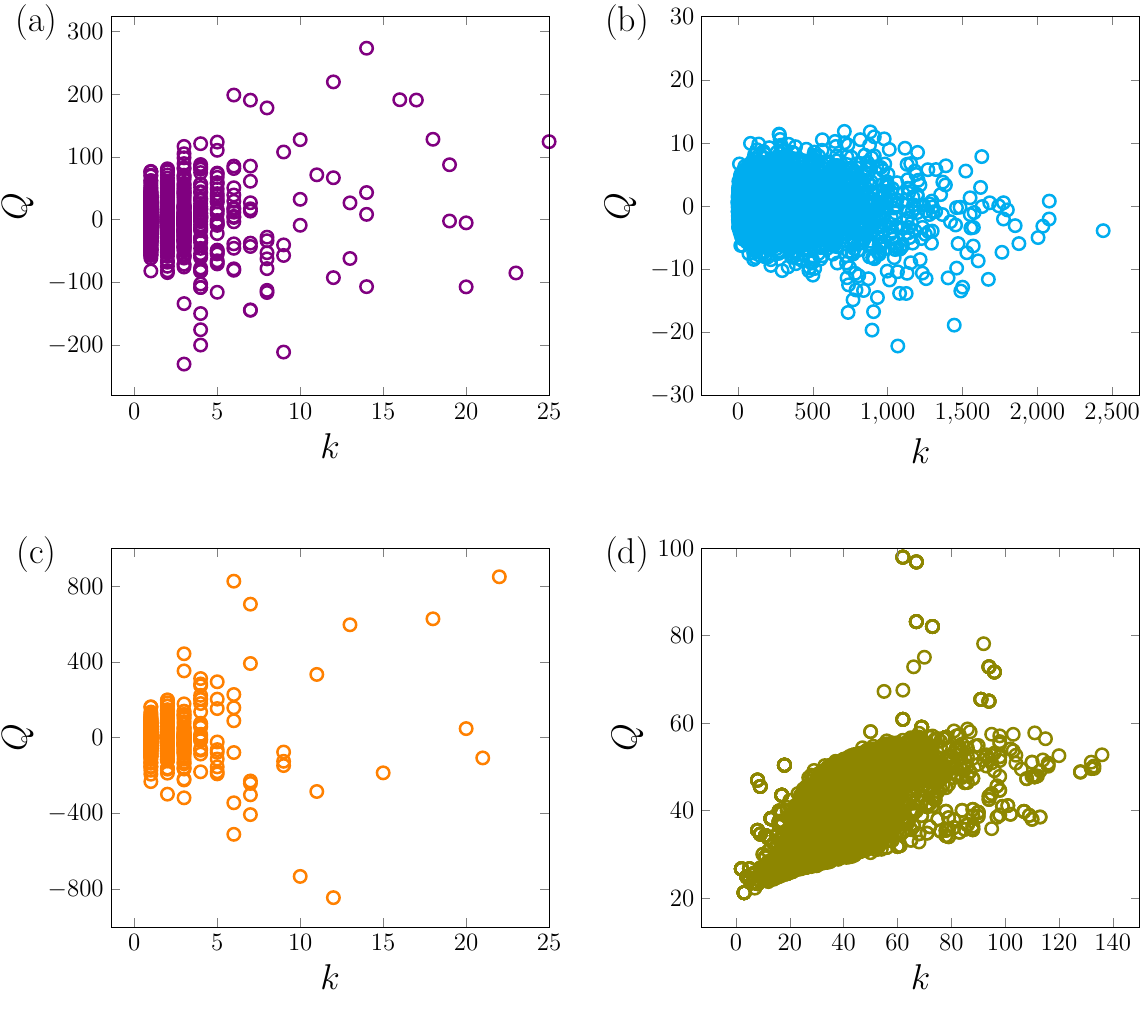}
    \caption{Size--charge correlations. Maps of cluster charge vs.\ cluster size at early (left) and late (right) times for the Cauchy--Lorentz case; the Gaussian case yields very similar results. The top row corresponds to neutral systems at $t=\SI{0.01}{\second}$ (early) and $t=\SI{220}{\second}$ (late), while the bottom row shows non-neutral systems at $t=\SI{0.098}{\second}$ (early) and $t=\SI{4e10}{\second}$ (late).
}

    \label{fig:fig5}
\end{figure*}

  In Fig.~\ref{fig:fig5} we show maps of cluster charge $Q$ versus cluster size $k$ at early times (left) and late times (right) for the Cauchy--Lorentz case; the Gaussian case exhibits a very similar behavior. At first glance, the differences between the two cases may appear modest; however, this is  due to the different axis scales, and a careful comparison reveals significant quantitative differences. As time progresses, the neutral system shows no discernible correlation between cluster charge and size, indicating that charge remains randomly distributed among clusters of all sizes. In contrast, the charged system evolves from an initially uncorrelated state—imposed by the initial conditions—to one in which a clear positive correlation develops between $q$ and $k$. 
At long times essentially all clusters acquire a positive charge, 
revealing that the net charge drives a systematic buildup of charge with cluster growth.

This correlation arises because collisions in a net–charged environment are not purely random: clusters that already carry more positive charge repel each other more strongly, reducing their collision frequency and slowing their neutralization by smaller, less charged aggregates. As aggregation proceeds, larger clusters accumulate a disproportionate fraction of the total charge through repeated mergers with like–charged partners, while repulsive interactions inhibit charge compensation. The result is a preferential charging of the largest clusters and a progressive drift of the entire population toward the sign of the net charge.

\section{Conclusions}
\label{SectV}

We have investigated the kinetics of charged–particle aggregation using Monte Carlo simulations of the Smoluchowski equation, focusing on the impact of different initial charge distributions on the temporal evolution 
of the system. In neutral systems, the long–time dynamics converge toward a universal SPD, confirming asymptotic
universality and the loss of memory of the initial conditions.
In contrast, net–charged systems do not reach a true self–preserving
regime; instead, electrostatic repulsion progressively suppresses
coagulation, leading to a charge–limited steady state characterized by a
finite mean cluster size and a quasi--stationary size distribution. However, for intermediate times, when the initial charges follow a Cauchy--Lorentz distribution, the largest cluster mass can exceed that of the Gaussian case by nearly a factor of twenty before converging to the same asymptotic state. 
This dramatic enhancement persists over a wide range of the initial charge distribution width and is largely independent of the net charge, demonstrating that the heavy tails of the Cauchy--Lorentz distribution strongly accelerate early-stage cluster growth.

Key differences also emerge between neutral and net–charged systems. In neutral suspensions, frequent collisions between oppositely charged clusters promote efficient charge exchange and drive the charge distribution toward a universal Gaussian form with no preferred cluster size. By contrast, when a net charge is present, electrostatic repulsion suppresses large–cluster collisions, producing a size distribution with a well–defined maximum and a charge distribution that drifts toward larger positive values while retaining memory of its initial width. These features reveal how a persistent net charge limits statistical mixing and sets a characteristic cluster scale, in sharp contrast to the scale–free growth of the neutral case.

These findings have direct implications for aerosol and dust coagulation. 
The intermediate time regime of rapid growth promoted by heavy-tailed charge statistics provides a natural mechanism to produce larger aggregates, which in turn are expected to have stronger electrostatic cohesive energy, and thus, more resistant to external perturbations. In addition, an initial heavy-tailed charge distribution could help to form larger clusters before other mechanisms can stop growth. We propose that this should be observed in other systems, in particular for larger size, athermal granular media. Although these are not driven by Brownian motion, the stronger likelihood of highly charged particles should also exist in tribocharged agglomerated oxide dust grains in a PPD. In this case, the faster formation of larger clusters could help bridge the gap between sub-millimeter dust and the pebble–scale seeds required for streaming instabilities and gravitational collapse in the process of planet formation~\cite{Yang2017}. 

Finally, it remains to be seen how the inclusion of fragmentation would modify the trends reported here. Fragmentation processes, which counterbalance coagulation by breaking large aggregates into smaller ones, have been shown to introduce steady-state regimes in related systems~\cite{fuhrer2025hybrid-d52}, in which cluster sizes and charges fluctuate around stationary values rather than growing indefinitely. In neutral systems, such a balance would likely produce symmetric, Gaussian‐like charge distributions with limited width, while in net-charged systems fragmentation could act as a charge–redistribution mechanism, mitigating the Coulomb bottleneck and maintaining a population of moderately charged clusters.
Another important direction is the inclusion of polarization effects. In the present study the collision kernel incorporates only Coulombian interactions, yet polarization forces—arising from charge–induced dipoles—are known to enhance attractive interactions, particularly at short range, and should therefore accelerate aggregation \cite{Cameron2025,Lee2018,Nakajima1999,Lindgren2018,Guo2021}. Polarization-driven charging and interactions have also been successfully modeled in previous studies \cite{yoshimatsu2017self}. 
Incorporating polarization would require modifying the kernel to account for charge–dipole and dipole–dipole contributions, which may substantially alter the growth dynamics and the resulting charge distributions.
Exploring both the aggregation–fragmentation interplay and the role of polarization forces represents a natural extension of the present work and may shed further light on the charge-regulated pathways leading to planetesimal formation, as well as other aggregation-related phenomena.

\begin{acknowledgments}
We thank the supercomputing infrastructure of the High-Performance Computing UOH laboratory of Universidad de O'Higgins, Rancagua. This work was funded by the Agencia Nacional de Investigación y Desarrollo (ANID) through the Fondecyt Grant 1221597. The authors acknowledge fruitful discussions with Sim\'on Casassus (DAS, U. de Chile), as well as with Andrea Stoellner and Scott R. Waitukaitis (ISTA).  
\end{acknowledgments}

\bibliography{references_new}

\end{document}